\documentclass[aps,prb,twocolumn,superscriptaddress,floatfix,longbibliography]{revtex4-2}
\usepackage{graphicx}
\usepackage{epstopdf}
\usepackage{upgreek}
\usepackage{amsmath}
\usepackage{amssymb}
\usepackage{color}
\usepackage{soul}
\usepackage[normalem]{ulem}
\usepackage{xr-hyper}
\usepackage[bookmarks=false,colorlinks,citecolor=blue]{hyperref}
\hypersetup{colorlinks=true,linkcolor=blue,filecolor=blue,citecolor = blue, urlcolor=blue}
\makeatletter
\newcommand*{\addFileDependency}[1]{
  \typeout{(#1)}
  \@addtofilelist{#1}
  \IfFileExists{#1}{}{\typeout{No file #1.}}
}
\makeatother
\newcommand*{\myexternaldocument}[1]{%
    \externaldocument{#1}%
    \addFileDependency{#1.tex}%
    \addFileDependency{#1.aux}%
}

\myexternaldocument{SI}
\newcommand{\pll}{\kern 0.2em/\kern -0.6em /\kern 0.2em}

\def\braket#1{\mathinner{\langle{#1}\rangle}}

\begin{document}
\title{Layer-Controlled Intermolecular Coupling and Many-Body Effects in C$_{60}$ Films}

\author{Hai-Lan Luo} \affiliation{Department of Physics, University of California, Berkeley, Berkeley, CA 94720, USA} \affiliation{Materials Sciences Division, Lawrence Berkeley National Laboratory, Berkeley, CA 94720, USA}
\author{Weitang Li} \affiliation{Guangdong Basic Research Center of Excellence for Aggregate Science, School of Science and Engineering, The Chinese University of Hong Kong, Shenzhen, Guangdong 518172, China}
\author{Luca Moreschini} \affiliation{Department of Physics, University of California, Berkeley, Berkeley, CA 94720, USA} \affiliation{Materials Sciences Division, Lawrence Berkeley National Laboratory, Berkeley, CA 94720, USA}
\author{Jonathan Denlinger} \affiliation{Advanced Light Source, Lawrence Berkeley National Laboratory, Berkeley, CA 94720, USA}
\author{Zhigang Shuai} \affiliation{Guangdong Basic Research Center of Excellence for Aggregate Science, School of Science and Engineering, The Chinese University of Hong Kong, Shenzhen, Guangdong 518172, China} \affiliation{MOE Key Laboratory for Organic OptoElectronics and Molecular Engineering, Department of Chemistry, Tsinghua University, Beijing 100084, China}
\author{Claudia Ojeda-Aristizabal} \affiliation{Department of Physics and Astronomy, California State University Long Beach, Long Beach, CA 90840, USA}
\author{Alessandra Lanzara} \email{alanzara@lbl.gov} \affiliation{Department of Physics, University of California, Berkeley, Berkeley, CA 94720, USA} \affiliation{Materials Sciences Division, Lawrence Berkeley National Laboratory, Berkeley, CA 94720, USA} \affiliation{Kavli Energy NanoScience Institute, University of California, Berkeley, Berkeley, CA 94720, USA}

\date{\today}

\begin{abstract}
\noindent {\bf Abstract}

Crystalline C$_{60}$ is a molecular solid whose electronic properties emerge from the interplay of intermolecular hopping, electron correlations, and electron-vibration coupling. Unlike moir$\rm\acute{e}$ van der Waals heterostructures, where interaction strength is commonly tuned by twist angle, molecular materials offer a complementary route in which layer number, molecular orientation, and substrate registry provide experimentally accessible control parameters. Here we present a systematic thickness-dependent angle-resolved photoemission study of C$_{60}$ films, spanning the monolayer to the bulk limit. The HOMO-derived band exhibits a non-monotonic evolution: the intermediate-thickness film shows larger bandwidth, reduced effective mass, and pronounced gap-like and sub-band features. The experimental trends, together with Holstein-model simulations, point to strengthened effective intermolecular electronic coupling and enhanced electron-phonon-induced spectral renormalization in the intermediate-thickness regime. These results identify a dimensional crossover in C$_{60}$ films and establish layer number as an effective knob for engineering electronic structure and many-body interactions in molecular thin films.


\end{abstract}

\maketitle

\noindent {\bf Introduction}

The ability to tune electronic structure and many-body interactions is central to the design of quantum materials\,\cite{DHsieh2017DBasov,JMoore2017BKeimer}. In van der Waals heterostructures, moir$\rm\acute{e}$ superlattices have demonstrated how structural control can reshape bandwidth, symmetry and interaction strength\,\cite{AMacDonald2011RBistritzer,PJarillo2018YCao}. Molecular solids, assemblies of molecules bound by van der Waals interactions, offer a complementary platform. Unlike conventional atomic crystals, they possess internal rotational and orientational degrees of freedom that directly influence intermolecular overlap and electronic coupling\,\cite{OGunnarsson1997,BPowell2011RMcKenzie,DJerome2004,TNematiaram2021ATroisi}. Their electronic properties are therefore governed not only by lattice symmetry, but also by intermolecular overlap, molecular orientation, and coupling to intramolecular vibrations. In molecular thin films, these structural degrees of freedom can be accessed through experimentally tunable parameters such as layer number, molecular orientation, and substrate registry, providing flexible knobs for tuning electronic structure and many-body interactions\,\cite{MCrommie2008YWang,VBrouet2004ZXShen,OGunnarsson1997}.

Among molecular crystals, crystalline C$_{60}$ stands out as a promising system, enabling diverse device functionalities\,\cite{GYu1995AHeeger,SGunes2007NSariciftci,HLi2012ZBao,KWojciechowski2015HSnaith,CAristizabal2017AZettl,ASaid2024SDeWolf} and allowing controlled studies of collective and correlated quantum phenomena\,\cite{PHeiney1991ASmith,FGugenberger1992HWuhl,AHebard1991AKortan,PAllemand1991JThompson,OGunnarsson1997,AGanin2010KPrassides}. Its electronic properties are intimately governed by competing interactions, including intermolecular hopping, electron-electron correlation, and electron-vibration coupling\,\cite{WYang2003,AGanin2008,JZhou2023,PAi2023ALanzara}. Disentangling these microscopic ingredients is therefore essential for understanding and ultimately controlling the emergent properties.

A natural route toward this goal is to tune these interactions through experimentally accessible structural parameters. In layered van der Waals materials, layer number has proven to be a powerful control parameter, revealing how interlayer coupling and symmetry shape electronic states\,\cite{EMcCann2006,BPartoens2006FPeeters,KMak2010THeinz1,TOhta2007ERotenberg,KMak2010THeinz2,ASplendiani2010FWang,YZhang2014ZXShen,AGupta2015SSeal}. This concept naturally extends to crystalline C$_{60}$ films, where changing layer number can modify dimensionality, intermolecular coupling, orientational order, and substrate-induced effects. Indeed, a variety of emergent states have been observed in C$_{60}$-based platforms across different dimensional realizations, such as superconductivity and magnetism in bulk compounds\,\cite{AHebard1991AKortan,PAllemand1991JThompson}, unconventional electronic and orientational phases in thin films\,\cite{LTjeng1991GSawatzky,JHou2001QZhu,WYang2003,VBrouet2004ZXShen,MCrommie2008YWang,PAi2023ALanzara}, and quantum transport in single-molecule devices\,\cite{HPark2000PMcEuen}. However, a systematic understanding of how dimensionality influences the microscopic electronic structure and many-body effects in pristine, well-controlled C$_{60}$ films remains lacking.

Building on established layer-by-layer growth of C$_{60}$ films and the ability of angle-resolved photoemission spectroscopy (ARPES) to resolve their dispersive band structure\,\cite{DLatzke2019ALanzara,NHaag2020BStadtmuller}, we systematically investigate the thickness-dependent electronic structure of C$_{60}$ films over a wide thickness range, from monolayer (1\,ML) to bulk-like thickness - 22 layers (22\,ML). By tracking the evolution of the highest occupied molecular orbital (HOMO) and its adjacent HOMO-1 band, we extract layer-dependent changes in bandwidth, dispersion, and effective mass. A pronounced crossover emerges in the intermediate-thickness regime, where the HOMO bandwidth is largest among the measured thicknesses, consistent with strengthened effective intermolecular electronic coupling. At the same thickness, gap-like features that are present across all thicknesses become most prominent, and a distinct sub-band appears within the HOMO. Together with Holstein-model simulations, these observations identify an intermediate-thickness regime in which intermolecular hopping and electron-vibration coupling are simultaneously enhanced, establishing layer number as an effective knob for tuning many-body interactions in molecular thin films.
\vspace{3mm}

\begin{figure*}[tbp]
\begin{center}
\includegraphics[width=1.5\columnwidth,angle=0]{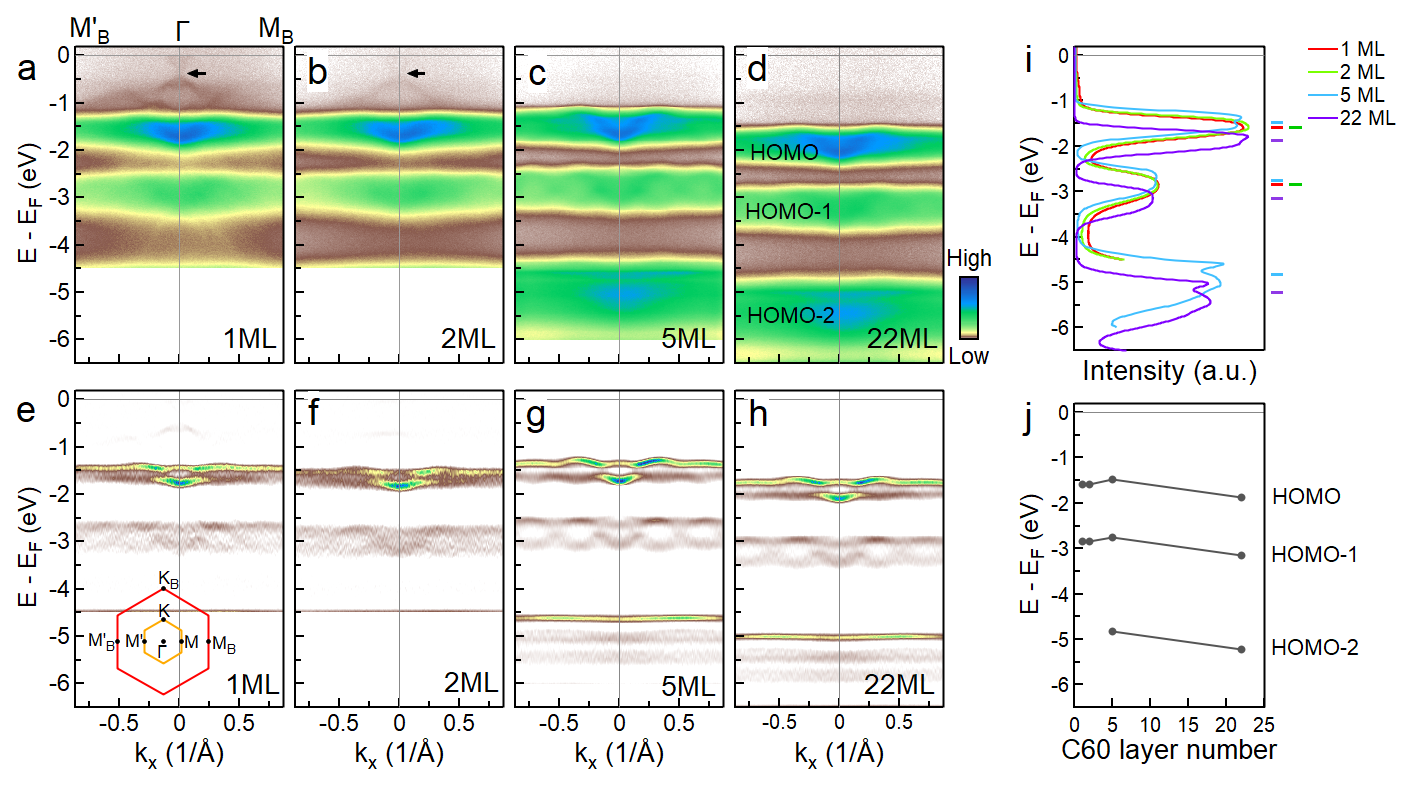}
\end{center}
\caption{\label{fig:fig1} \textbf{Thickness-dependent electronic structures of C$_{60}$ films on a Bi$_2$Se$_3$ substrate.} {\textbf{a-d,}} ARPES band structures of C$_{60}$ films with thicknesses of 1\,ML\,(a), 2\,ML\,(b), 5\,ML\,(c), and 22\,ML\,(d) C$_{60}$, measured along the M$\rm{_B}'$-$\Gamma$-M$\rm{_B}$ high-symmetry direction. Bands originating from the Bi$_2$Se$_3$ substrate are indicated by arrows in (a,b). {\textbf{e-h,}} Second-derivative images with respect to energy, derived from (a)-(d). The Brillouin zones of C$_{60}$ and Bi$_2$Se$_3$ are outlined in yellow and red, respectively, in the inset of (e). {\textbf{i,}} EDCs integrated over the full M$\rm{_B}'$-$\Gamma$-M$\rm{_B}$ momentum range for all four thicknesses. The colored bars on the right indicate the centroid energy of each band manifold. {\textbf{j,}} Thickness dependence of the centroid energies of each band manifold, extracted from the EDCs.
}
\end{figure*}

\noindent {\bf Results}

Taking advantage of the structural adaptability of C$_{60}$ molecules to various substrates, we grow C$_{60}$ thin films crystallizing in the $\it P$$\bar a$$\it{3}$ structure, with thickness ranging from 1\,ML to bulk-like regime, by molecular beam epitaxy (MBE) on the (111) surface of the topological insulator Bi$_2$Se$_3$. Due to the 3.4$\%$ uniaxial compressive strain imposed by lattice mismatch, the rotational disorder of C$_{60}$ molecules is suppressed, facilitating the resolution of dispersive molecular orbitals in ARPES\,\cite{DLatzke2019ALanzara}. The suppression of rotational disorder stabilizes a well-defined intermolecular overlap, enabling the thickness-dependent evolution of electronic coupling observed here.
This controlled platform enables a systematic investigation of how layer number modulates intermolecular coupling and electronic structure. 
Fig.~\ref{fig:fig1} presents the evolution of the electronic structure of C$_{60}$ as a function of film thickness along the $\rm{M}'_B$-$\Gamma$-$\rm{M}_B$ high-symmetry direction. We focus on the two uppermost valence bands, which are mainly derived from the highest occupied molecular orbital (HOMO) and the next-highest occupied molecular orbital (HOMO-1). The highly dispersive character of these bands is more clearly visualized in the second-derivative images (Fig.~\ref{fig:fig1}e-h), with the band extrema located at $\Gamma$, M and M$'$. 

Residual band features from the Bi$_2$Se$_3$ substrate are also visible in the thinner films (1 and 2\,ML), as indicated by arrows in Fig.~\ref{fig:fig1}a,b, but disappear in the thicker ones (5 and 22\,ML; Fig.~\ref{fig:fig1}c,d) because of the surface sensitivity of ARPES. This evolution reflects the progressive decoupling of the C$_{60}$ electronic states from the substrate with increasing thickness.

To further examine the thickness dependence of the valence states, Fig.\ref{fig:fig1}i displays the integrated energy distribution curves (EDCs) for the four film thicknesses. The full width at half maximum (FWHM) of the HOMO and HOMO-1 peaks remains nearly constant at 500\,meV and 670\,meV, respectively, across all thicknesses. This indicates that the overall quasiparticles lifetime and disorder broadening remain comparable across the series. In addition, the centroid energies of all bands exhibit a non-monotonic dependence on film thickness (Fig.~\ref{fig:fig1}j). While their positions remain nearly constant in the 1-5\,ML regime, a pronounced downward shift of approximately 300\,meV is observed in the 22\,ML film. This shift likely reflects charging effects associated with  low conductivity and wide band gap of thick C$_{60}$ layers.
\vspace{3mm}

\begin{figure*}[tbp]
\begin{center}
\includegraphics[width=1.5\columnwidth,angle=0]{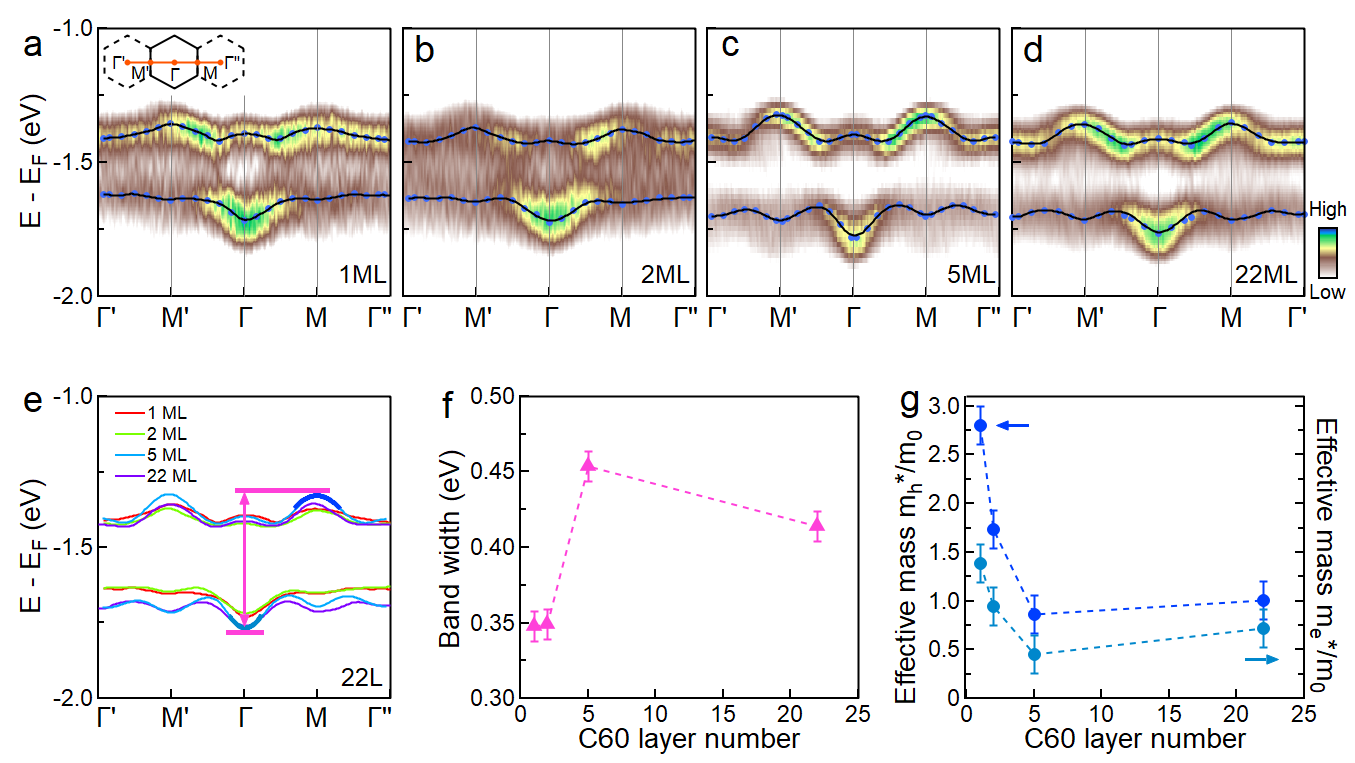}
\end{center}
\caption{\label{fig:fig2} \textbf{Thickness-dependent band dispersion and effective mass of HOMO band.} {\textbf{a-d,}} Second-derivative ARPES spectra of the HOMO band along the M$'$-$\Gamma$-M high-symmetry direction for C$_{60}$ films with thicknesses of 1\,ML\,(a), 2\,ML\,(b), 5\,ML\,(c), and 22\,ML\,(d). Blue dots indicate extracted peak positions of the dominant HOMO-derived dispersive spectral envelope, and black curves represent fits to these features. {\textbf{e,}} Fitted dispersions for all thicknesses overlaid for comparison. Blue curves show parabolic fits around the band top at M and the band bottom at $\Gamma$ for the 22\,ML film. The magenta arrow indicates the HOMO bandwidth ($\it W$) of the dominant HOMO-derived dispersive spectral envelope, defined as the energy difference between the fitted band maximum at M/M$'$ and the fitted band minimum at $\Gamma$. {\textbf{f,}} Thickness dependence of $\it W$. {\textbf{g,}} Effective hole masses ($\it m_h^{\ast}$) at M/M$'$ and electron masses ($\it m_e^{\ast}$) at $\Gamma$, extracted from local parabolic fits near the corresponding band extrema.
}
\end{figure*}

A quantitative analysis of the band dispersion is presented in Fig.~\ref{fig:fig2}. Panels a-d show enlarged second-derivative images of the HOMO band along the M$'$-$\Gamma$-M high-symmetry direction for C$_{60}$ films with four thicknesses. Two dispersive branches are clearly resolved by fitting the data (black curves in Fig.~\ref{fig:fig2}a-d): an upper branch with its band maximum at the M/M$'$ points and a lower branch with its minimum at $\Gamma$. The fitted dispersions for all thicknesses are overlaid in Fig.~\ref{fig:fig2}e for comparison. From these results, we extract the HOMO bandwidth ($\it W$) and effective masses ($\it m^{\ast}$) as a function of film thickness, as plotted in Fig.~\ref{fig:fig2}f,\,g, respectively. The HOMO-derived valence states are associated with the fivefold-degenerate molecular h$_u$ HOMO manifold of C$_{60}$, which can give rise to multiple HOMO-derived bands in the crystalline film\cite{MGolden1995ESohmen}. Given the multi-band nature of this manifold and the matrix-element sensitivity of C$_{60}$ photoemission\cite{DLatzke2019ALanzara2}, a unique subband-by-subband assignment is not feasible from the present ARPES data alone. We therefore define the HOMO bandwidth $\it W$ as the experimentally extracted energy range of the dominant HOMO-derived dispersive spectral envelope along the measured M$'$-$\Gamma$-M direction, rather than as the bandwidth of a single isolated tight-binding band.

The extracted HOMO bandwidth and effective masses reveal a clear dependence on film thickness. The HOMO bandwidth $\it W$ is approximately 350\,meV in 1\,ML and 2\,ML C$_{60}$ films, increases to $\sim$\,450\,meV in the 5\,ML film, and then decreases by $\sim$\,40\,meV in the much thicker 22\,ML film (Fig.~\ref{fig:fig2}f). In molecular solids, the HOMO bandwidth is strongly influenced by intermolecular orbital overlap and transfer integrals. However, because the HOMO-derived valence states of crystalline C$_{60}$ form a multi-band manifold, the extracted $\it W$ should not be interpreted as a quantitative measure of a unique microscopic hopping parameter. This non-monotonic behavior indicates that layer number acts as an effective tuning parameter for intermolecular hopping, with a maximum coupling reached at intermediate thickness. 
A similar thickness dependence is observed in the local band curvature.
By performing parabolic fits to the HOMO band near the band top at M and the band bottom at $\Gamma$, we extract the effective masses of holes ($\it m_h^{\ast}$) and electrons ($\it m_e^{\ast}$) as functions of layer number (Fig.~\ref{fig:fig2}g). Both $\it m_h^{\ast}$ and $\it m_e^{\ast}$ reach their minimum values at 5\,ML and increase significantly towards the monolayer limit. The reduced effective mass in the 5\,ML film is consistent with a more dispersive HOMO-derived band and larger bandwidth among the measured thicknesses. We also note that although the 1 and 2\,ML films exhibit comparable HOMO bandwidths, their extracted local effective masses differ by approximately a factor of two. This difference is not unexpected because the HOMO-derived states form a multi-band manifold rather than a simple single-band system governed by a single hopping parameter. Therefore, the HOMO bandwidth and local effective mass should be regarded as complementary empirical indicators of the HOMO-derived dispersion, rather than as two equivalent measures of a unique microscopic hopping parameter. Together, these results identify an intermediate-thickness regime characterized by a larger measured HOMO bandwidth and reduced local effective masses, consistent with strengthened effective intermolecular electronic coupling.
\vspace{5mm}

\begin{figure*}[tbp]
\begin{center}
\includegraphics[width=1.7\columnwidth,angle=0]{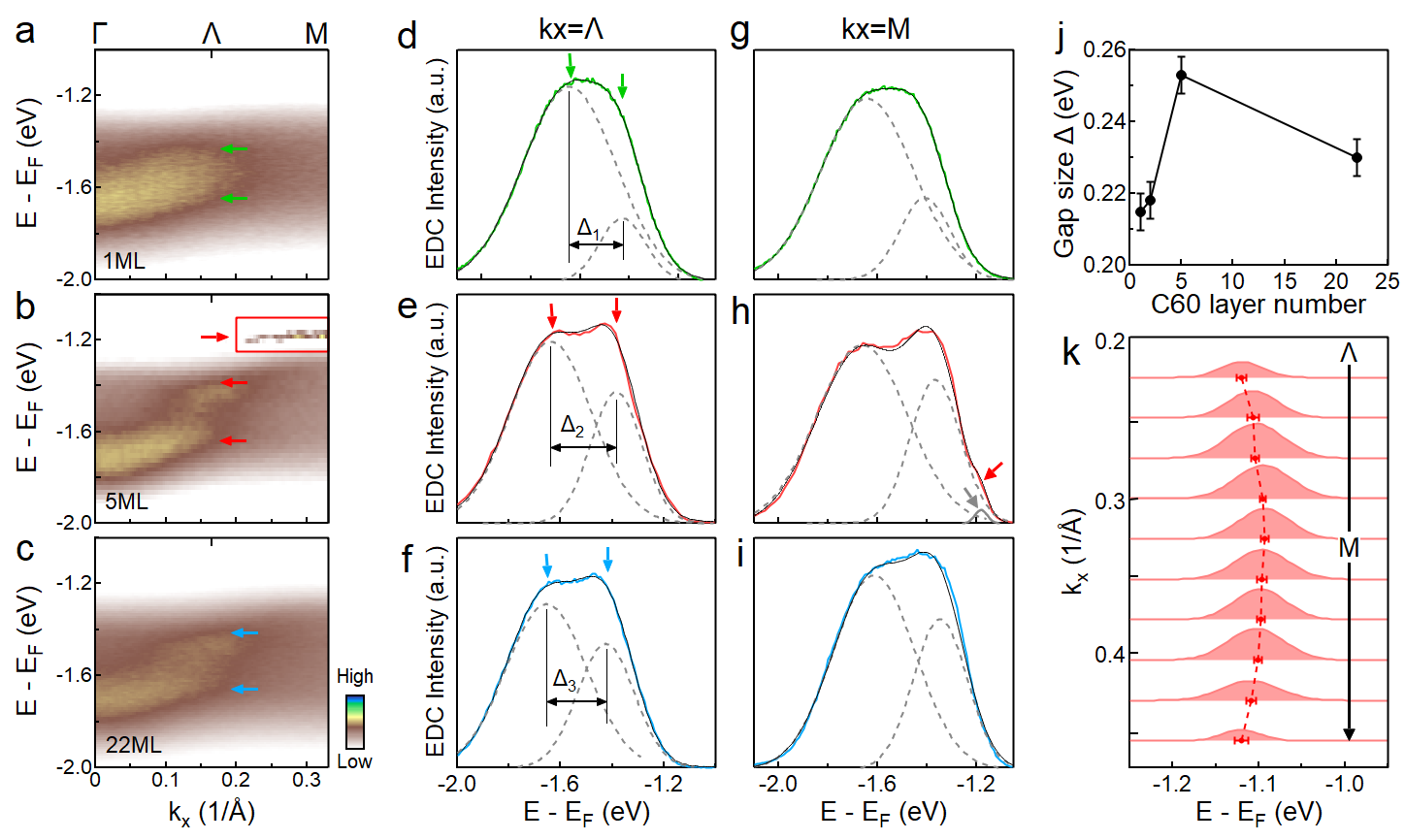}
\end{center}
\caption{\label{fig:fig3} \textbf{Thickness-dependent gap-like and sub-band features.} {\textbf{a-c,}} Raw ARPES intensity maps of the HOMO band dispersion for 1\,ML\,(a), 5\,ML\,(b), and 22\,ML\,(c) C$_{60}$ films. The $\Lambda$ point denotes the midpoint between $\Gamma$ and M. The inset in (b) shows the second-derivative image, highlighting the sub-band feature in the 5\,ML film; the corresponding raw-intensity evidence is provided in Supplementary Fig. S3. {\textbf{d-f,}} EDCs at the $\Lambda$ point for C$_{60}$ films with thicknesses of 1\,ML\,(d), 5\,ML\,(e), and 22\,ML\,(f), showing gap-like structures. Spectra are fitted with two Gaussian components, as shown by gray dashed curves. {\textbf{g-i,}} EDCs at the M point for C$_{60}$ films with thicknesses of 1\,ML\,(g), 5\,ML\,(h), and 22\,ML\,(i). A shoulder feature is observed in the 5\,ML film and is fitted with an additional Gaussian component, as indicated by arrows in (h). {\textbf{j,}} Extracted gap size at the $\Lambda$ point as a function of thickness. {\textbf{k,}} Momentum-dependence of the sub-band feature in the 5\,ML film along the $\Lambda$-M direction, showing a hole-like dispersion. 
}
\end{figure*}

Notably, the non-monotonic evolution of bandwidth and effective mass is accompanied by other thickness-dependent spectral reconstructions: a gap-like feature appears at the $\Lambda$ point in all films but is most pronounced at 5 ML, where a distinct sub-band feature also emerges. The coincidence of these features with the maximum bandwidth suggests a common underlying interaction driving both band dispersion and spectral renormalization. Fig.~\ref{fig:fig3}a-c present the raw data of the HOMO band of representative 1\,ML, 5\,ML, and 22\,ML films; the 2\,ML data are omitted because it closely resemble that of the 1\,ML film. 

To quantify the thickness-dependent evolution of the gap-like feature, we compare the EDCs at the $\Lambda$ point (Fig.~\ref{fig:fig3}c-f). Each EDC can be well reproduced by two Gaussian components (gray dashed curves), clearly indicating the presence of a gap. The extracted gap sizes, $\Delta_1$, $\Delta_2$, and $\Delta_3$, for the 1\,ML, 5\,ML and 22\,ML films, respectively, evolve non-monotonically with thickness and peak at 5\,ML, as summarized in Fig.~\ref{fig:fig3}j. This behavior mirrors the evolution of the bandwidth, suggesting a correlation between effective intermolecular coupling and the strength of the spectral gap.

In addition to the gap-like feature at $\Lambda$, the EDCs at the M point also exhibit a notable thickness dependence. As shown in Fig.~\ref{fig:fig3}g-i, a pronounced shoulder near the band top is observed only in the 5\,ML C$_{60}$ film, and is absent in the 1\,ML and 22\,ML films. To quantify this feature, we fit the EDCs at M. The spectra of the 1\,ML and 22\,ML films (Fig.~\ref{fig:fig3}g and \ref{fig:fig3}i) are well described by two Gaussian components (gray dashed curves), yielding excellent overall fits (black curves). In contrast, the EDC of the 5\,ML film (Fig.~\ref{fig:fig3}h) cannot be satisfactorily described using only two Gaussian components; a third component (gray arrow) is required to account for the shoulder on the low-binding-energy side (red arrow). This additional spectral component indicates the emergence of a new electronic feature that is stabilized only at intermediate thickness. 

Momentum-resolved analysis of this sub-band (Fig.~\ref{fig:fig3}k) reveals a hole-like dispersion with a band maximum at M, closely resembling the main HOMO band. Its spectral weight, however, is highly localized in momentum space and decays rapidly beyond $k_{\rm{M}} \pm 0.12\,\text{\AA}^{-1}$. A parabolic fit to this sub-band yields an energy separation of $\sim$\,180\,meV from the HOMO maximum. The similarity in dispersion together with the fixed energy offset suggests that this feature does not originate from a distinct band, but rather from a many-body renormalization of the primary electronic state.

The gap-like feature and sub-band formation are common spectroscopic signatures of electron-phonon coupling, as reported across semiconductors\,\cite{MKang2018KKim,CChen2018MAsensio}, superconductors\,\cite{JLee2014ZXShen} and organic molecular materials\,\cite{GGoodvin2006GSawatzky,SKera2009NUeno,SCuichi2011SFratini,WLi2021ZShuai}. In weakly coupled molecular films, where the bandwidth ($W$) is comparable to the intramolecular vibrational energy ($\Omega$), electron-vibration coupling can produce shake-off replica bands below the main band\,\cite{SKera2009NUeno}. The energy separation between the main band and its replicas is determined by $\Omega$, while their relative intensities reflect the coupling strength. In molecular crystals with stronger intermolecular coupling ($\it{W}\,\textgreater\,\rm\Omega$), electron-phonon interactions can give rise to more complex spectral features. Specifically, theoretical studies have predicted that the HOMO band can split into a series of sub-bands separated by gaps\,\cite{GGoodvin2006GSawatzky,SCuichi2011SFratini,WLi2021ZShuai}. These considerations suggest that the observed thickness-dependent spectral features may originate from a coupling between electronic states and intramolecular vibrational modes. 
\vspace{5mm}

\begin{figure*}[tbp]
\begin{center}
\includegraphics[width=1.7\columnwidth,angle=0]{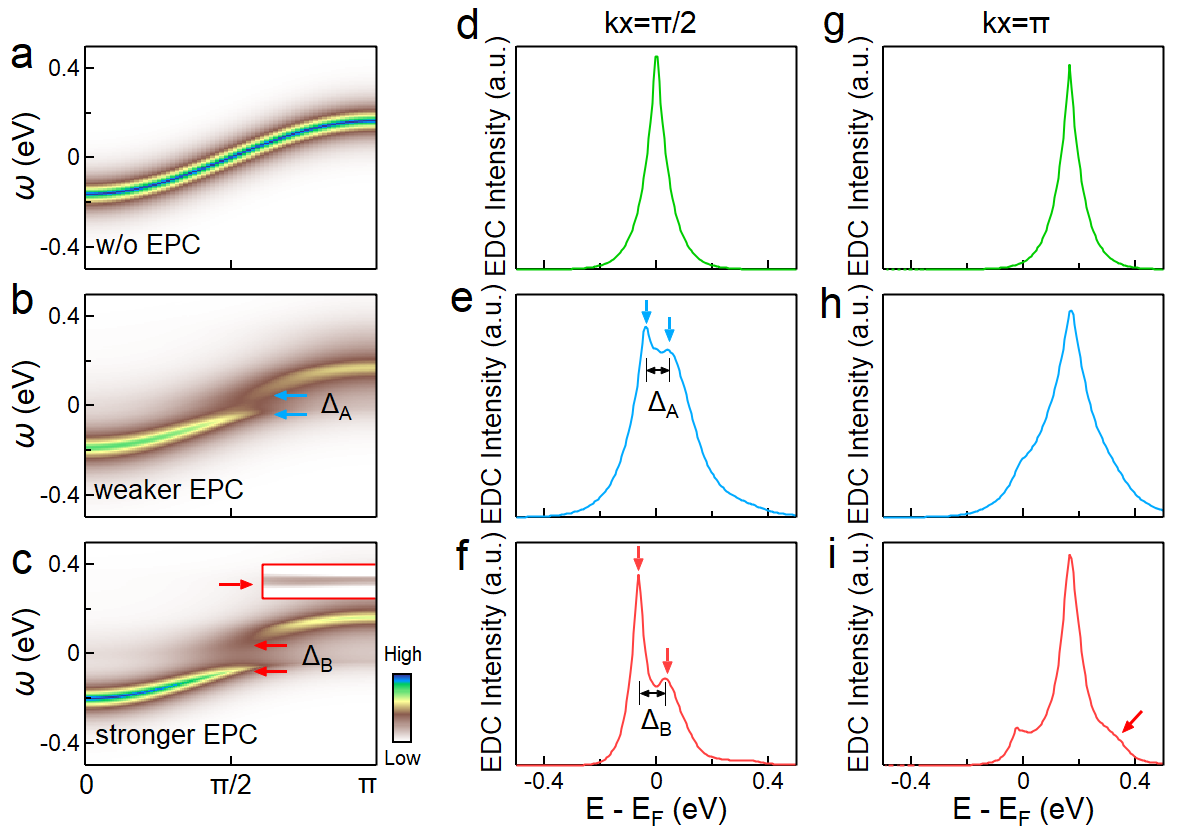}
\end{center}
\caption{\label{fig:fig4} \textbf{Simulated spectral functions with varying electron-phonon coupling strength.} {\textbf{a-c,}} Simulated single-particle spectral functions of the HOMO band based on a Holstein model with coupling to the  H$\rm{_g}$(7) (1443\,cm$^{-1}$) and H$\rm{_g}$(8) (1607\,cm$^{-1}$) intramolecular phonon modes. The mode-resolved reorganization energy ($g_m^2\omega_m$) for each of these two modes is set to 0 meV in (a), 20 meV in (b), and 30 meV in (c), corresponding to the no-coupling, intermediate-coupling, and stronger-coupling cases, respectively. Arrows indicate gap-like features and sub-band formation. {\textbf{d-f,}} Corresponding EDCs at $k_{\pll}=\pi/2$ for the three electron-phonon coupling strengths, showing the evolution from a single peak to a split structure with increasing coupling strength. {\textbf{g-i,}} Corresponding EDCs at $k_{\pll}=\pi$ for the same three electron-phonon coupling strengths, showing the emergence of a sub-band feature at stronger coupling.
}
\end{figure*}

To test this hypothesis, we identify the phonon modes most likely to contribute to the coupling. In typical systems, the energy separation between the main band and its replica is set by the energy of the coupled phonon mode\,\cite{SKera2009NUeno,JLee2014ZXShen,MKang2018KKim,CChen2018MAsensio}. In the 5\,ML C$_{60}$ film, the observed energy separation of 180\,meV closely matches two high-frequency intramolecular phonon modes, H$\rm{_g}$(7) and H$\rm{_g}$(8), with vibrational wavenumbers of 1443\,cm$^{-1}$ (178\,meV) and 1607\,cm$^{-1}$ (198\,meV), respectively\,\cite{OGunnarsson1995,ZHuang2020DLiu}. Moreover, among all H$\rm{_g}$ phonons in C$_{60}$, these two modes exhibit the strongest electron-phonon coupling via the Jahn-Teller effect\,\cite{ZHuang2020DLiu}. The quantitative agreement between the sub-band separation and the vibrational energies provides strong evidence that these modes dominate the observed spectral renormalization. We therefore incorporate H$\mathrm{_g}$(7) and H$\mathrm{_g}$(8) into our simulations.

We emphasize that this Holstein-model calculation is not intended as a full multi-band fit to the experimental HOMO manifold. Instead, it provides a minimal phenomenological description to test whether coupling to high-frequency intramolecular modes can generate gap-like features and sub-band spectral weight comparable to those observed experimentally. To enable comparison with our experimental data, we calculate the single particle spectral function using the Holstein model, both without and with electron-vibration coupling. Fig.~\ref{fig:fig4}a shows the simulated spectral function described by a tight-binding dispersion, $\omega$($\it k_{\pll}$)\,=\,$\omega_0$\,-\,2$\it V$cos($\it {ak}_{\pll}$), with a Lorentzian broadening of 50\,meV. Here, $\omega_0$ is set to 0\,meV as the energy reference, and the effective hopping parameter $\it V$\,=\,87.5\,meV is chosen to reproduce the approximate energy scale of the monolayer HOMO bandwidth $\it W$ (Fig.~\ref{fig:fig2}). The simulated EDCs at $k_{\pll}=\pi/2$ and $\pi$, taken from Fig.~\ref{fig:fig4}a, both display single-peak structures (Figs.~\ref{fig:fig4}d,\,g), as expected in the absence of electron-phonon coupling. This baseline calculation establishes that the additional spectral features observed experimentally do not arise from the underlying band structure alone. 

Figure~\ref{fig:fig4}b shows the simulated spectral function in the presence of electron-phonon interactions. The H$\rm{_g}$(7) and H$\rm{_g}$(8) modes are included with equal coupling strengths, corresponding to a reorganization energy of 20\,meV (see Methods). The resulting spectrum exhibits a gap ($\Delta_{\mathrm{A}}$) at $k_{\pll}=\pi/2$, but no sub-band feature near $\pi$, as confirmed by the corresponding EDCs obtained at $\pi/2$ (Fig.~\ref{fig:fig4}e) and $\pi$ (Fig.~\ref{fig:fig4}h). 
This intermediate-coupling regime reproduces the partial spectral renormalization observed experimentally, characterized by the presence of a gap without a well-defined sub-band. In particular, it captures the behavior of the thicker (22ML) film, where spectral reconstruction is present but remains limited in strength.

In contrast, Fig.~\ref{fig:fig4}c presents the spectral function calculated with stronger electron-phonon interactions. Here, both H$\rm{_g}$(7) and H$\rm{_g}$(8) modes are assigned larger coupling strengths, corresponding to a reorganization energy of 30\,meV. A clear gap opens near $k_{\pll}=\pi/2$, and a sub-band emerges above the HOMO top at $\pi$, as highlighted by the second-derivative image. Quantitatively, the EDC at $\pi$/2 in Fig.~\ref{fig:fig4}f clearly displays a distinct two-peak structure, with a gap size $\Delta_{\mathrm{B}}$ larger than $\Delta_{\mathrm{A}}$, reflecting the effect of enhanced electron-phonon coupling. Meanwhile, the EDC at $\pi$ in Fig.~\ref{fig:fig4}i shows a shoulder feature associated with the sub-band above the HOMO top, as indicated by an arrow. Importantly, the energy separation between the sub-band and the HOMO top agrees well with the experimental observations and is consistent with the characteristic vibrational energy scale identified experimentally. This strong-coupling regime simultaneously captures both the enlarged gap and the emergence of the sub-band, providing a unified description of the spectral features observed at intermediate thickness. In particular, it reproduces the behavior of the 5 ML film, where spectral renormalization is most pronounced within the measured thickness series, linking the observed electronic reconstruction to an enhanced electron-phonon interaction strength. Notably, our simulations show that the number and spacing of phonon-induced gap-like features are governed primarily by the energy scale of the dominant phonon modes, with high-frequency H$\rm{_g}$(7)- and H$\rm{_g}$(8)-dominated coupling naturally yielding a single prominent gap opening rather than multiple closely spaced gaps, as discussed in the Supplementary Note and shown in Fig. S2.

\vspace{5mm}

\noindent {\bf Discussion}

The thickness-dependent evolution of the HOMO bandwidth and local effective mass in C$_{60}$ films suggests strengthened effective intermolecular electronic coupling in the intermediate-thickness regime. Rather than evolving monotonically, these quantities reach extrema in the intermediate-thickness regime, followed by possible saturation or weakening in thicker films, identifying layer number as an effective control parameter for tuning electronic dispersion in molecular solids. Strikingly, this trend closely parallels the evolution of the gap-like feature and the emergence of the sub-band, which are also most pronounced at the same thickness. Although charging may affect the precise quantitative value extracted for the 22\,ML bandwidth, the overall conclusion does not rely on this quantity alone. The 5\,ML film, as the measured representative of the intermediate-thickness regime, simultaneously exhibits enhanced HOMO-derived dispersion, reduced local effective masses, the most pronounced gap-like feature, and the clearest sub-band spectral weight. It is unlikely that charging alone would produce such a correlated evolution across several independent observables. Therefore, while the precise quantitative values extracted for the 22\,ML film should be interpreted with caution, the overall trend toward strengthened effective intermolecular electronic coupling and enhanced electron-phonon-induced spectral renormalization in the intermediate-thickness regime remains robust. The energy separation between the main HOMO band and the sub-band in the 5\,ML film hints at the involvement of the high-frequency intramolecular vibrational modes, H$\rm{_g}$(7) and H$\rm{_g}$(8). By incorporating these modes into a Holstein model, we reproduce the experimentally observed thickness dependence of both the gap-like feature and the sub-band under varying electron-phonon coupling strengths. Taken together, these results demonstrate that layer thickness simultaneously tunes both intermolecular coupling and electron-phonon coupling, linking band dispersion and many-body renormalization through a single experimentally accessible control parameter.

To understand the origin of this crossover, we examine the role of strain. A uniaxial compressive strain of 3.4$\%$ has been reported for C$_{60}$ films grown on Bi$_2$Se$_3$\,\cite{DLatzke2019ALanzara}, and this strain is expected to relax as the film thickness increases. Compressive strain reduces intermolecular spacing and enhances wavefunction overlap, thereby naturally accounting for the stronger intermolecular hopping in the 5\,ML film than in the 22\,ML film\,\cite{Nigam2012RRao,ABakulin2015DCahen,TKubo2016JTakeya}. However, strain alone cannot explain the weaker coupling in 1-2\,ML films, which are expected to experience compressive strain no smaller than that in the top layer of the 5\,ML film. A likely origin of this behavior is static disorder from the Bi$_2$Se$_3$ substrate. In this scenario, the bottom one or two C$_{60}$ layers are more susceptible to electrostatic potential fluctuations caused by surface defects, such as Se or Bi vacancies\,\cite{CMann2013CShih}, which can generate trap states and hinder coherent intermolecular hopping\,\cite{WKalb2010BBatlogg}. 
The observed crossover can therefore be understood as the result of competing effects: substrate-induced disorder suppresses coherence in the ultrathin limit, while strain relaxation reduces intermolecular overlap in thicker films. In addition to these effects, thickness-dependent changes in screening and molecular orientation may also contribute to the observed HOMO bandwidth and local band curvature. The intermediate thickness regime balances these effects, leading to enhanced electronic coherence and interaction strength.

The origin of the thickness dependence of electron-phonon coupling likely involves multiple factors. One plausible contribution is strain, which has been shown to modulate electron-phonon coupling strength in some quantum materials\,\cite{CSi2013FLiu}. Whether this mechanism also applies to C$_{60}$ films remains an open question. A second possibility is a thickness-dependent interlayer effect. In layered systems with strong interlayer coupling, film thickness can directly influence electron-phonon coupling strength\,\cite{JHe2022RZhang}; although C$_{60}$ is generally regarded as a molecular solid with weak interlayer van der Waals interactions, related effects cannot be ruled out a priori. Charge transfer from the substrate may also play a role by enhancing interlayer effects through induced dipoles\,\cite{KOhno2001YKawazoe}, although direct evidence for such transfer in C$_{60}$/Bi$_2$Se$_3$ under our experimental conditions is currently lacking\,\cite{RPandeya2024AGruneis}. Taken together, these considerations suggest 
that the thickness dependence of electron-phonon coupling arises from an interplay of strain, disorder, and interlayer effects. While the relative contributions of these mechanisms remain to be fully resolved, our results demonstrate that layer thickness provides an effective experimental knob for tuning electron-phonon interactions in molecular thin films. 

In particular, the simultaneous enhancement of the gap-like feature and sub-band at intermediate thickness establishes a direct link between layer-controlled electronic structure and electron-phonon-induced spectral renormalization.
While previous experiments on another molecular system, rubrene, have identified the presence of a gap-like feature arising from electron-phonon coupling\,\cite{FBussolotti2017SKera}, the systematic evolution of band renormalization with coupling strength has remained largely unexplored. Our results provide such a thickness-resolved perspective, revealing how electron-phonon coupling evolves across the crossover from the two-dimensional to the three-dimensional limit.

More broadly, these findings may also shed light on quantum phenomena related to electron-phonon coupling in C$_{60}$, including the role of high frequency intramolecular phonons in unconventional superconductivity\,\cite{Dresselhaus1996,JZhou2023LYang}. Beyond the specific case of C$_{60}$, our results highlight molecular thin films as a versatile platform for engineering electronic structure and many-body interactions. In addition to dimensionality, molecular orientation provides an additional tuning axis for intermolecular coupling, distinguishing molecular solids from conventional layered materials. In contrast to moir$\rm\acute{e}$-based approaches in van der Waals materials, where interaction strength is tuned through twist angle and heterostructure stacking, molecular systems offer a complementary route based on layer number, molecular orientation, and substrate registry. This flexibility provides new opportunities for designing tunable quantum states in molecular solids using experimentally accessible structural parameters.
\vspace{4mm}

\noindent {\bf Methods}
\vspace{1mm}

\noindent {\bf Molecular-beam epitaxy growth of C$_{60}$ films on Bi$_2$Se$_3$.} C$_{60}$ films are grown $\it{in}$ $\it{situ}$ on freshly cleaved Bi$_2$Se$_3$ (001) surfaces, following the procedure described in our previous work\,\cite{DLatzke2019ALanzara}. The 2 ML, 5 ML and 22 ML C$_{60}$ films are measured in the $\it {as}$-$\it{grown}$ state, while the 1 ML C$_{60}$ film is obtained by annealing a thicker film. To obtain the 1 ML C$_{60}$ film, we first grow a $\sim$5 ML C$_{60}$ film on Bi$_2$Se$_3$, and then anneal the sample at $\sim$220$^\circ$C, leaving a monolayer on the Bi$_2$Se$_3$ surface. The low-energy electron diffraction (LEED) pattern of the annealed monolayer C$_{60}$ film confirms that it is well-ordered with a hexagonal lattice structure, as shown in the Supplementary Materials of Ref.\,\cite{DLatzke2019ALanzara}). This growth approach enables precise control of film thickness, providing a platform to systematically investigate layer-dependent electronic structure.
\vspace{1mm}

\noindent {\bf High resolution ARPES measurements.} High-resolution angle-resolved photoemission spectroscopy (ARPES) measurements were performed at Beamline 4.0.3 (MERLIN) of the Advanced Light Source. The data were acquired using 45\,eV linearly $p$\,-\,polarized light, which is predominantly out of the sample plane. The energy resolution was set to $\sim$20\,meV, with an angular resolution of $\sim$0.2 $^\circ$. All data were taken at 20\,K under a base pressure better than 5$\times$10$^{-11}$\,Torr. These conditions allow direct resolution of the dispersive molecular orbitals and thickness-dependent spectral features discussed in the main text.
\vspace{1mm}

\noindent {\bf Single-particle spectral function with electron-phonon coupling.} The momentum-resolved single-particle spectral function is calculated as: 
\begin{equation}
    A(k, \omega) = \frac{1}{N\pi} \sum^N_{mn} e^{ikR(m-n)} \int_0^\infty \braket{c_m(t) c_n^\dagger(0)} e^{i \omega t} dt 
\end{equation}
using the time-dependent density matrix renormalization group (TD-DMRG) method\,\cite{WLi2021ZShuai, JRen2022ZShuai}.
The time evolution is performed based on the time-dependent variational principle.
The C$_{60}$ molecules are modeled through a one-dimensional multi-mode Holstein Hamiltonian
\begin{equation}
\label{eq:ham}
\begin{aligned}
    \hat H & = \hat H_{e} + \hat H_{ph} + \hat H_{e-ph} \\ 
    \hat H_e & = V \sum_n (c_{n+1}^\dagger c_n +  c_{n}^\dagger c_{n+1}) \\
    \hat H_{ph} & = \sum_{n, m} \omega_m b^\dagger_{n, m} b_{n, m} \\
    \hat H_{e-ph} & = \sum_{n, m} g_m \omega_m (b^\dagger_{n, m} + b_{n, m}) c^\dagger_n c_n
\end{aligned}
\end{equation}
where $c^\dagger$ ($c$) and $b^\dagger$ ($b$) are the creation (annihilation) operators for electrons and phonons, respectively. $V$ denotes an effective hopping parameter in the minimal one-dimensional Holstein model. Its value is chosen to reproduce the approximate energy scale of the monolayer HOMO bandwidth observed in Fig.~\ref{fig:fig2}, but it should not be regarded as a unique microscopic hopping parameter of the multi-band C$_{60}$ HOMO-derived valence states. $\omega_m$ is the frequency of the $m$\,th normal mode, and $g_m$ is the dimensionless electron-phonon coupling constant between the $m$\,th mode and the HOMO orbital of C$_{60}$. \textit{Ab initio} calculations have shown that the two high-frequency intramolecular modes, H$\rm{_g}$(7) and H$\rm{_g}$(8), dominate the total reorganization energy, defined as $\lambda=\sum_m g^2_m \omega_m$\,\cite{ZHuang2020DLiu}, which serves as a measure of the electron-phonon coupling strength. Our experimental results also suggest an important role of high-frequency phonons. Accordingly, we set $\omega_1$\,=\,1443\,$\textrm{cm}^{-1}$ (H$\rm{_g}$(7)) and $\omega_2$\,= 1607\,$\textrm{cm}^{-1}$ (H$\rm{_g}$(8)) in our simulations to capture their contribution to the electron-phonon coupling.  

Their corresponding reorganization energies, $g_m^2\omega_m$, are set to 0\,meV (Fig.\,~\ref{fig:fig4}a), 20\,meV (Fig.\,~\ref{fig:fig4}b), and 30\,meV (Fig.\,~\ref{fig:fig4}c) for each phonon mode. These values are chosen to span the weak- to strong-coupling regimes discussed in the main text. To account for the effects of low-frequency modes and environmental dissipation, we apply a Lorentzian broadening $\frac{t^2}{\gamma^2+t^2}$ to the correlation function $\braket{c_m(t) c_n^\dagger(0)}$, where $\gamma$ is chosen such that the sum of the total reorganization energy and $\gamma$ equals 110 meV.
\vspace{3mm}

\noindent {\bf Supporting Information}\\
Note on phonon-induced gap formation; background-subtracted EDC analysis; simulated spectral functions for different electron-phonon coupling distributions; and raw-intensity evidence for the weak sub-band feature in the 5\,ML C$_{60}$ film.
\vspace{3mm}

\noindent {\bf Acknowledgments}\\
We thank Drew W. Latzke for experimental help, and Jack Broad and Sin$\rm\acute{e}$ad M. Griffin for helpful discussions. This work was primarily supported by the U.S. Department of Energy, Office of Science, Office of Basic Energy Sciences, Materials Sciences and Engineering Division, under contract No. DE-AC02-05CH11231 within the van der Waals heterostructure Program (KCWF16). This research used resources of the Advanced Light Source, a US DOE Office of Science User Facility, under Contract No. DE-AC02-05CH11231. C.O.-A. acknowledges the U.S. Department of Energy, Office of Science, Office of Basic Energy under award number DE-SC0018154 for film growth and ARPES data acquisition.
\vspace{3mm}

\noindent {\bf Author Contributions}\\
A.L. conceived the project. J.D. and C.O.-A. contributed to the film growth and performed the ARPES experiments. H.L.L. analyzed the ARPES data with help from L.M. W.L. and Z.S. contributed to the calculations. H.L.L. and A.L. wrote the manuscript with input from all co-authors. All authors participated in discussions and commented on the manuscript.
\vspace{3mm}

\renewcommand{\bibsection}{%
  \par\noindent\textbf{References}\par\medskip
}

\bibliographystyle{achemso}
\bibliography{Ref}

\end{document}